\documentclass[conference]{IEEEtran}
\IEEEoverridecommandlockouts
\usepackage{cite}
\usepackage{amsmath,amssymb,amsfonts}
\usepackage{algorithmic}
\usepackage{graphicx}
\usepackage{textcomp}
\usepackage{xcolor}
\def\BibTeX{{\rm B\kern-.05em{\sc i\kern-.025em b}\kern-.08em
    T\kern-.1667em\lower.7ex\hbox{E}\kern-.125emX}}
\begin{document}

\title{Security and Privacy Challenges in Deep Learning Models
}

\author{
\IEEEauthorblockN{Gopichandh Golla}
\IEEEauthorblockA{\textit{Computer Science and Engineering} \\
\textit{The State University of New York at Buffalo}\\
Buffalo, USA \\
ggolla@buffalo.edu}
}

\maketitle

\begin{abstract}
These days, deep learning models have achieved great success in multiple fields, right from autonomous driving to medical diagnosis. These models have expanded the abilities of artificial intelligence by offering great solutions to complex problems that were very difficult to solve earlier.  In spite of their unseen success in various, it has been identified, through research conducted, that deep learning models can be subjected to various attacks that compromise model security and data privacy of the Deep Neural Network models. Deep learning models can be subjected to various attacks at different stages of their lifecycle. During the testing phase, attackers can exploit vulnerabilities through different kinds of attacks such as Model Extraction Attacks, Model Inversion attacks, and Adversarial attacks. Model Extraction Attacks are aimed at reverse-engineering a trained deep learning model, with the primary objective of revealing its architecture and parameters. Model inversion attacks aim to compromise the privacy of the data used in the Deep learning model. These attacks are done to compromise the confidentiality of the model by going through the sensitive training data from the model's predictions. By analyzing the model's responses, attackers aim to reconstruct sensitive information. In this way, the model's data privacy is compromised. Adversarial attacks, mainly employed on computer vision models, are made to corrupt models into confidently making incorrect predictions through malicious testing data. These attacks introduce and change the input data in such a way that it is not recognizable as adversary data, causing deep learning models to mis-classify and make wrong decisions. Models can be subjected to attacks not only evaluation phase but also during the training phase. Data Poisoning Attacks introduce malicious examples into the training dataset, which makes the training phase ineffective and leads to the loss of the integrity of the deep learning model. Data privacy can be compromised by not only model inversion attacks but also by service providers of AI models.

This Survey is done to gain an in-depth understanding of the security and privacy problems of Deep neural nets. It is done by analyzing what kinds of attacks are present, how each type of attack works, and its challenges and drawbacks. This paper also discusses what are the potential evolutions of these attacks moving forward as well.

\end{abstract}


\section{Introduction}
These days, deep learning models have achieved great success in multiple fields, right from medical diagnosis and natural language processing to autonomous driving and image classification. In healthcare, Deep learning models are utilized in disease prediction and medical image analysis with greater accuracy. In finance, they are involved in fraud detection systems. Computer Vision models of Deep learning help self-driving cars perform in complex environments in terms of navigation. Natural Language Processing models are improving virtual assistants and language translation tools. The major reason behind the success of Deep neural net models is their great ability to learn patterns from vast datasets available. Advances in computing power and advanced algorithm innovation have also assisted in the success of Deep neural network models. On the flip side, integration of these advanced models into a wide range of domains has come with its own set of security and privacy vulnerabilities that require thorough investigation and defense techniques.

The vulnerabilities of these Deep neural net models pose severe challenges and threats to the integrity, reliability, and confidentiality of these systems. A compromised DL model can lead to bad decision-making, improper working of autonomous systems, or breaches of confidential data. These vulnerabilities may lead to financial losses to threats to public safety[2]. For example, a compromised Autonomous driving system may lead to accidents and loss of lives. Face recognition systems used in biometric authentication systems can be compromised by introducing noise to the system thus compromising the Integrity of the systems.

This survey is conducted to analyze and understand the length and breadth of attacks on Deep Learning systems. Attacks are categorized into 4 types, such as Model Extraction Attacks, Model Inversion attacks, Adversarial attacks, and Poisoning Attacks[1]. This survey discusses different techniques and their challenges and drawbacks along with the potential evolution of the attacks in the future. 

Model Extraction and Model inversion attacks are aimed at compromising the privacy of the DL models and exposing model parameters and training data. Model Extraction attack is aimed at duplicating the DL model's parameters and architecture. Whereas, Model Inversion attacks are aimed at reproducing confidential information by going through already available data. Adversarial attacks( Evasion attacks) and poisoning attacks' main purpose is to influence prediction results either by providing falsified testing data and getting wrong predictions with confidence or obstructing the model training during the training phase with the help of adversarial examples. Evasion attacks are done during the evaluation phase and are further classified into white box and black box attacks. beforehand knowledge of the model is necessary in order to execute white box attacks where attackers use these data to form perturbed inputs to cause undesirable outcomes. Gradient-based techniques are the common mode of operations employed to perform white-box attacks. Black box attacks are more common between these 2 types as this resembles real-world scenarios much better where the attackers don't have prior knowledge of the models. Queries are made to understand the model's behavior to disturb the performance of the models. Here, just the model's behavior is known, but not the model's architecture or hyperparameters. Poisoning attacks are done during the training phase of the model by introducing wrongful data into the trained dataset. These kinds of attacks are under-researched and are divided into 3 subcategories, they are performance degradation, targeted poisoning, and backdoor attacks. In consideration of the impact of adversarial threats, extensive research has been dedicated to developing defense mechanisms against adversarial attacks. Techniques such as adversarial training, gradient masking, and robust optimization aim to bolster DL models' resilience to adversarial perturbations. Adversarial attacks are not constrained to a single domain, they can affect a wide range of domains or architectures such as CNNs, RNNs, etc.[3]

To summarize, This Survey is done to gain an in-depth understanding of the security and privacy problems of Deep neural nets. It is done by analyzing what kinds of attacks are present, how each type of attack works, and its challenges and drawbacks. This paper also discusses what are the potential evolutions of these attacks moving forward as well.

\section{Main Techniques}

\subsection{Model Extraction Attacks}

In this type of attack, the attacker doesn't know the internal workings of the model and the aim is to duplicate the model. The attacker queries the model with input x to get the result y. By performing extensive querying as mentioned before, the attacker gains enough information needed to reverse engineer the model and be able to replicate its internal working. This information can be further used to perform subsequent attacks like adversarial attacks on the model. The adversary model for this attack typically operates in a black-box manner, where the attackers have no information on how the model works and no information regarding the architecture, parameters, hyperparameters, and training dataset is known. The attacker can only interact with the model through prediction APIs. Moreover, they do not have access to data that matches the distribution of the target model's training data. There are also limitations on query frequency, as attackers may face restrictions or be blocked by the API if they submit queries too frequently.

Model extraction can be done in 3 different ways, such as Equation solving, Metamodel training, and Training substitute model. In equation solving, a typical classification machine learning model that works on continuous function can be solved by having enough pairs of input and output values. In the metamodel training approach(classifier of classifiers), attackers query the model to gain the output for a specific input. These values are used to train a metamodel to replicate the original model, thus used later to predict the original model's parameters. Training substitute model is similar to the metamodel training approach where a set of input-output value pairs are trained on the substitute model and that model's parameters mimic the original model's parameters.

\subsection{Model Inversion Attacks}

Model Inversion attack is a kind of attack employed to gain information about the training data of Neural Networks models. Model Inversion attack uses a concept called inverse information flow. Since neural networks can remember a lot of information, inverse information flow can be used to gain insights into data properties, thus compromising the data privacy of Deep learning models. Model Inversion attack can be done in both white box and black box settings. Model Inversion attack can be done in 2 different ways, such as Membership Inversion attack and property inversion attack. The presence of specific data records in the training dataset can be determined using a Membership Inversion attack, whereas Statistical property inference can be achieved through a Property Inference Attack.

\subsubsection{\textit{Membership Inference Attack :}}

Membership inference attack is a kind of Model inversion attack whose main aim is to determine the presence of specific data in the training dataset or not. This attack can be done in both black-box and white-box setups. The working of this kind of attack is done in 4 steps, namely, Data synthesis, Shadow model training, attack model training, and membership determination. In the data synthesis step, data can be generated either manually or by training generated models. As part of shadow model training, shadow models are trained to generate training data for the attack model. The attack model is trained using the data generated above to determine whether a given instance of the training dataset is part of the training dataset or not is done attack model training step. In the membership determination step, the determination of whether the input instance is part of the training dataset or not.

\subsubsection{\textit{Property Inference Attack :}}

A property Inference Attack is a kind of Model Inversion Attack where the main goal of the attack is to determine the characteristics of the dataset. A workflow of this kind is similar to a Membership inference attack but the end goal is different. Here, the dataset is divided based on the presence of a certain feature.

\subsection{Adversarial Attacks}

This kind of attack is also known as an evasion attack where adversarial examples are applied during the inference phase(testing phase), thereby causing the model to incorrectly classify the input data. This attack can be done on all kinds of models such as image classification, speech recognition, text processing, and malware detection. Adversary examples used are not recognizable by humans but can deceive the target models. Adversarial attacks can also be done in both white-box and black-box settings. Adversarial attacks are efficiently developed in image classification models. Here, the distance(perturbation) between the adversarial example and the true example should be as minimal as possible. This distance can be measured using Lp norms.
Here the attacks can be either targeted or non-targeted. 

\subsubsection{\textit{White-box attacks :}}

In this type of attack, attackers already have intrinsic information about the target models. perturbations required to generate adversarial examples can be obtained by either calculating gradients or solving optimization problems. The first attack that we discuss here is done using the Limited-memory Broyden-Fletcher-Goldfarb-Shanno( L-BFGS) algorithm to generate adversarial examples[4]. Similar to the above attack, CW-Attack minimizes a different objective function to obtain perturbations[5]. Jacobian-based Saliency Map Attack ( JSMA ) finds the perturbations by calculating the saliency maps of gradients derived from forward propagation[6]. This saliency map is then used to find the perturbations. L0 is used as a distance metric here. Fast Gradient Sign Method ( FGSM ) : FGSM explores the linearity of neural net models to drive the input data towards the gradients to obtain needed perturbations[7]. This model can be helpful in linearly related simple models. Iterative FGSM: Unlike the traditional FGSM model, the iterative FGSM model computes the perturbated examples step by step by applying the FGSM multiple times[8]. This method is faster than the traditional model with higher success rates as well.

\subsubsection{\textit{Black-box attacks :}}

In general, black box attacks are closer to the real world as model details, such as architecture, parameters, hyperparameters, and gradients, are not known to attackers most of the time. Industrial and mobile deployed models are kept confidential and are not accessible to adversaries. The following are some of the Black-box attacks.

Transfer-based attacks[4] are developed based on the assumption of inherent linearity of the neural nets, allowing a substitute model to approximately predict the decision boundary of the actual model. This substitute model's architecture is determined based on the input format, such as images or text. Zeroth-order optimization is a popular form of attack where attackers approximate the gradient using only the Model's prediction API. However, these are computationally expensive to implement. In Limited Information attacks, Chen et al.[9] proposed a method that approximates the gradient of a target model using only the output scores thus minimizing the amount of information needed. Suet et al.[10] introduced a technique where only one pixel is changed to produce adversarial examples. In Predicted Label-based attacks, Only predicted labels are used to produce the adversary examples due to limited information availability. Boundary attacks use random walks to approach the decision boundary iteratively.

\subsection{Poisoning Attacks}

These kinds of attacks are employed during the training phase of the target model by introducing malicious examples to weaken the model training. These kinds of attacks are under-researched to date and are classified into 3 categories, namely performance degradation attacks, targeted poisoning attacks, and backdoor attacks.

\subsubsection{\textit{Performance degradation attack:} }
In this kind of attack, performance degradation is achieved using malicious examples that are developed using bi-level optimization.

\subsubsection{\textit{Targeted Poisoning Attack:}}
 The main aim of this attack is to misclassify certain test examples during testing by modifying training data based on changes to parameters and loss functions.

\subsubsection{\textit{Backdoor attack:}}
 In this attack, the main aim is to create a backdoor to the model that can be exploited during the testing phase. Insertion of subtle triggers like patches or watermarks into training samples and forcing neural nets to produce false outputs.

\section{Issues and Problems}
Extracting machine learning models using approaches like equation solving, training metamodels, and training substitute models contains certain challenges and drawbacks. Solving equations is not so complex for small models. For larger models, equations become more complex thus leading to less practicality of solving them. Equation solving cannot be a good choice when quick extraction of models is necessary. The equation-solving method is limited to classification models that deal with continuous class variables. When it comes to training metamodels and training substitute models, large amounts of data are required. So, querying the models to get relevant data can be a challenging process. In addition to collecting large amounts of data, the computation of substitute and metamodel models requires considerable amounts of computational resources. Substitute models trained cannot perform well on complex target models or models that are trained on diverse data. 

With respect to Model Inversion attacks, the Shadow model training step can be a time-consuming task and require a good amount of computational resources. One of the challenges with respect to data synthesis is the step collection of training data that closely replicates the training data can be really difficult if there is no extensive knowledge about the dataset's specifics is available. The effectiveness of the models developed cannot be consistent across all the attacks and can be ineffective at times when the type of data is diverse or the target model is more complex. One of the main drawbacks of model inversion attacks is the amount of resources required to train shadow models is expensive. Strongly defended target models cannot be attacked with this kind of attack

White box attacks contain their own set of challenges and drawbacks. L-BFGS attack is computationally expensive and has a high misclassification rate. CW attack also has its challenges such as tuning hyperparameters and solving the complex objective function is computationally expensive. One of the challenges of JSMA attack is the calculation of accurate salient maps and one of its major drawbacks is its high reliance on the L0 norm to calculate distances between the adversary and true examples. Since Iterative FGSM depends on calculating gradient values iteratively, convergence is not guaranteed in all scenarios and may be tiresome. Determining the Optimal number of iterations required can be a challenge.

One of the challenges in black-box attacks is a limited amount of information with respect to the target model is available. Substitute model creation is really expensive on both data collection and model training fronts. Model APIs block the users when they query frequently, thus posing a challenge to collect the data required. Drawbacks of black-box attacks consist of vulnerability to defenses mainly substitute models. Not all attacks are effective. Models that are developed to attack a target system can be effective on only that model, but not on other models thus limiting the model's transferability.

For poisoning attacks, developing perturbed examples can be a challenging task when attackers have a limited amount of training data. This kind of attack can also require intensive resources. One of the challenging tasks in these kinds of attacks is the creation of good adversarial examples that look like actual samples. These kinds of attacks are limited to models to which they are aiming to perturb, and cannot be transferred to other target models.

\section{Future trends}

Deep learning is a discipline that is always evolving, and there is a never-ending battle between attackers and defenders over model security. 

\subsection{\textit{Model Extraction Attacks :}}

Attackers may create "advanced query strategies" that query models to obtain information with fewer queries, which is one possible progression of model extraction attacks. To obtain more accuracy, attackers can also combine various extraction model combinations. One evolution in model extraction attacks is improvement in transfer learning.

\subsection{\textit{Model Inversion Attacks :}}

These days, model inversion attacks on complex systems are not as effective; nevertheless, in the future, with the development of sophisticated methodologies, complicated models may be inverted. As security continues to advance, attackers can leverage the most recent knowledge to circumvent advanced data privacy strategies. 

\subsection{\textit{Adversarial Attacks :}}

One possible development in adversarial attacks is the ability to expand beyond attacks on a certain target model to additional models. This guarantees that transfer learning will improve. This development of more complicated adversarial attacks may target multiple different models at the same time. In addition, adversaries might create algorithms that operate in real time and drastically reduce time overhead. 

\subsection{\textit{Poisoining Attacks :}}

In terms of poisoning attacks, altered instances produced during the model's training phase will resemble real examples even more and become more difficult to identify during the training phase of the models. In the future, backdoor attacks may develop into more sophisticated ones that may include more sophisticated trigger mechanisms and stealthy activation functions. Attackers may utilize recent advancements in the field of reinforcement learning and develop even better backdoor triggers. Attackers may modify their poisoning tactics in response to the target model's defense measures.

In summary, It's important to note that as attacks evolve, the field of AI security will also evolve to develop better defenses and strategies to mitigate these threats.







\end{document}